\documentclass[12pt]{article}

\usepackage{fullpage}
\usepackage{latexsym}

\newcommand{\be}{\begin{equation}}
\newcommand{\ee}{\end{equation}}
\newcommand{\ba}{\begin{eqnarray}}
\newcommand{\ea}{\end{eqnarray}}
\newcommand{\bd}{\begin{displaymath}}
\newcommand{\ed}{\end{displaymath}}

\def\thalf{{\textstyle{\frac{1}{2}}}}

\def\rootg{\sqrt{-g}}

\def\DL{\mathcal{D}_L}

\begin{document}
\title{
{\bf Sound Mode Hydrodynamics from Bulk Scalar Fields}}

\author{{T. Springer} \vspace*{0.1in}\\
{\it School of Physics and Astronomy}\\
{\it University of Minnesota}\\
{\it Minneapolis, Minnesota 55455, USA}}

\date{May 22, 2009}

\maketitle

\begin{abstract}
We study the hydrodynamic sound mode using gauge/gravity
correspondence by examining a generic black brane background's
response to perturbations.  We assume that the background is generated
by a single scalar field, and then generalize to the case of multiple
scalar fields.  The relevant differential equations obeyed by the
gauge invariant variables are presented in both cases.  Finally, we
present an analytical solution to these equations in a special case;
this solution allows us to determine the speed of sound and bulk
viscosity for certain special metrics.  These results may be useful in
determining sound mode transport coefficients in phenomenologically
motivated holographic models of strongly coupled systems.
\end{abstract}

\newpage
\section{Introduction}
The plasma created in heavy ion collisions at RHIC appears to be both
strongly coupled, and well described by hydrodynamics \cite{whitepapers, Molnar, Huovinen}.
  Hydrodynamics is an effective theory which describes the
behavior of a fluid on length and time scales which are much longer
than any microscopic scale.  The hydrodynamic stress-energy tensor is
constructed from conserved quantities, and is built so that it
respects equilibrium thermodynamics and the symmetries in the problem.
One can then add plane wave type perturbations $\sim e^{i(q z - w 
t)}$, and use the conservation equation $\partial_\mu T^{\mu \nu}$ to
analyze the normal modes of the system - see
\cite{hydroreview,Kovtun2} for reviews.  For the shear mode, the
dispersion relation between the energy of the perturbation $w$ and
the momentum $q$ is found to be
\be
	w(q)_{shear} = -i \frac{\eta}{\epsilon + P} q^2,
	\label{sheardispersion}
\ee
while for the sound mode it is
\be
	w(q)_{sound} = v_s q - i \frac{\eta}{\epsilon + P} \left( \frac{p-1}{p} + \frac{\zeta}{2\eta} \right)q^2.
	\label{sounddispersion}
\ee
Here, $p$ is the number of spatial dimensions, $\epsilon$ and $P$
denote the equilibrium pressure and energy density, respectively,
$v_s$ is the speed of sound and $ \eta$ and $\zeta$ are the shear viscosity and
bulk viscosity.  It is desirable to calculate these latter three quantities
which are used in hydrodynamic models of heavy ion collisions, but in
the regime of strong coupling, perturbative calculations are
unreliable.

The AdS/CFT correspondence \cite{Maldacena, Witten, Gubser}, (or more
generally, the `gauge/gravity duality' or `holography') has become an
indispensable tool for describing strongly coupled systems, and has
enjoyed arguably its greatest success calculating hydrodynamic
transport coefficients.  Within this framework, hydrodynamic
dispersion relations are calculated by computing correlation functions
of the stress-energy tensor, and then either examining the poles of
such quantities, or by using Kubo relations \cite{Policastro1,
recipe,hydroI, hydroII, Herzog1, Herzog2}.  Alternatively, one can
examine behavior of a background under perturbations, and determine
the dispersion relation by imposing appropriate boundary conditions
\cite{Kovtun1, Mas}.  The resulting dispersion relations from all of
these methods coincide as demonstrated in \cite{Kovtun1}.  After
computing the dispersion relation, one can compare with the formulas
(\ref{sheardispersion}), (\ref{sounddispersion}), and read off the
appropriate transport coefficients.

Another method which has been employed to access such quantities is
the `membrane paradigm', in which one computes the hydrodynamic
properties of a black hole's `stretched horizon'.  These hydrodynamic
quantities agree with the AdS/CFT calculations in all cases tested so
far \cite{stretched, Saremi, Kapusta, Natsuume, Liu}.  The details of
the membrane paradigm approach are not relevant for this paper, but we
mention these works for historical reasons.  In \cite{stretched}, a
formula for the shear viscosity to entropy density ratio $\eta/s$ was
derived which is applicable to a wide variety of gravity duals.
Application of this formula to known gravity duals always resulted in
$\eta/s =1/4 \pi$, and eventually led to the celebrated viscosity
bound conjecture that $\eta/s \geq 1/4 \pi$ for all physical
substances.  While the membrane paradigm approach was first used to
derive this general formula, it can also be derived by examining the
quasi-normal mode spectrum as shown in \cite{StarinetsMembrane}.
Clearly, a similar general formula for the bulk viscosity would also
be desirable in the hopes that other universal behavior might be
discovered.  Such a formula would also be useful when attempting to
phenomenologically fit the results from the lattice \cite{Kharzeev1, Kharzeev2}.

This work is motivated by the goal of developing such a formula,
though such a formula will not be presented here.  Instead,
this work should be viewed as a first step towards this goal.  Here,
we consider sound mode fluctuations of a dual gravity theory supported
by a single scalar field, and then generalize to the case of multiple
scalar fields.  We attempt to be as general as possible by
not specifying the background profiles of the scalar field(s).

The recent works \cite{Gubser1,Gubser2} also explore the sound mode
in phenomenologically motivated single scalar models.  The work presented
here is complementary to these papers, though we allow for the 
possibility of multiple scalar fields, and also do not restrict ourselves
to five dimensions.   

In section 1, we write the relevant equations that must be obeyed by
the background fields.  In section 2, we introduce the standard sound
mode perturbations and explicitly compute the linearized Einstein
equations.  In section 3, we reduce these Einstein equations to
two gauge invariant equations.  In section 4, we illustrate how one
uses these equations to solve for the hydrodynamical dispersion
relation and thus compute the speed of sound and bulk viscosity.  The
gauge invariant equations are rather complicated, so we only consider
a simple special case in this section.  The results of this section
are generalizations of the results of \cite{Mas}.  Finally, in
section 5, we generalize our gauge invariant equations by including an
arbitrary number of scalar fields.  We conclude and mention some
prospects for further investigation in section 6.

Our general relativistic conventions are those of \cite{Weinberg}.  
we use $R_{\mu \nu}, G_{\mu \nu}$ and $T_{\mu \nu}$ to denote
the Ricci, Einstein, and stress-energy tensors respectively.  We 
will also make use of the shorthand notation
\ba
	R^0_{\,0} &\equiv& F_0 \nonumber \\
	R^x_{\,x} &\equiv& F_x \nonumber \\
	R^r_{\,r} &\equiv& F_r.
\ea
In Appendix \ref{ricciappendix}, one can find explicit forms of these functions in
terms of the metric components.  $\nabla_{\mu}$ denotes the 
covariant derivative, and $\Box \equiv \nabla_\mu \nabla^\mu$.  
Throughout the paper, we also use the notation $\DL$ to denote the 
logarithmic derivative, namely
\be
	\DL \left[X(r)\right] \equiv \frac{X'(r)}{X(r)}.
	\label{DL}
\ee

\section{Background Fields}
We wish to examine hydrodynamic fluctuations on the following $p+2$
dimensional gravitational background
\be
	ds^2 = g_{00}(r) dt^2 + g_{xx}(r) dx_j dx^j + g_{rr}(r) dr^2
\ee
where $j = 1,2...p$, and $r$ is the extra-dimensional coordinate. We will often make
use of the definition
\be
	f(r)  \equiv  \sqrt{-g_{00}(r)g^{xx}(r)}. \label{fdef}\\
\ee

We assume the position of a horizon at $r = r_0$, and that the behavior
of the metric components near the horizon is
\ba
	g_{00}(r) &\approx& -\gamma_0 (r-r_0) + \mathcal{O}(r-r_0)^2 \nonumber\\
	g_{rr}(r) &\approx& \frac{\gamma_r}{r-r_0} + \mathcal{O}(1) \nonumber \\
	g_{xx}(r) &\approx& g_{xx}(r_0).
\label{NH}
\ea
The quantities $\gamma_0, \gamma_r$ and $g_{xx}(r_0)$ are independent of $r$.  The Hawking 
temperature is given by
\be
	T = \frac{1}{4\pi}\sqrt{\frac{\gamma_0}{\gamma_r}}.
\ee

Let us make the simplest assumption, that this
metric is created by a single, minimally coupled scalar field.  In
other words, we assume the action is of the form
\be
	\mathcal{S} = \frac{1}{16 \pi G_{p+2}} \int \, d^{p+2}x \rootg 
	\left( R  - \thalf \partial_\mu \phi \partial^\mu \phi - U(\phi) \right). 
\ee
Then, the background equations are
\ba
	G_{\mu \nu} &=& -8 \pi G_{p+2} T_{\mu \nu} \\
	\Box \phi &=& \frac{dU}{d\phi}.
\ea
Here we ignore any subtleties regarding boundary terms that come from
integration by parts.  Such terms can be taken care of by adding
additional boundary terms to the action, though we have not explicitly
written such terms above.

The energy-momentum tensor derived from the action is
\ba
	8 \pi G_{p+2} T_{\mu \nu} &=& \thalf \left(\partial_\mu \phi \partial_\nu \phi - g_{\mu \nu} \mathcal{L}_\phi \right) \\
	\mathcal{L}_\phi &=& \thalf \partial_\lambda \phi \partial^\lambda \phi + U(\phi)
\ea
Because the background metric only depends on the extra dimensional
coordinate $r$, it is clear that the field $\phi$ is also only a
function of $r$.  In the case of multiple scalar fields, it might be
possible to have fields which depend on the other coordinates,
provided that all such dependence cancels out in the combination that
appears in $T_{\mu \nu}$.  We will not consider such special cases,
and will always assume that the scalar fields only depend on $r$.

In this case, we can get a more explicit form of the potential:
\be 
	\label{Uprime}
	\frac{dU(\phi(r))}{d\phi} = \Box \phi = \frac{1}{\rootg} \partial_r \left[\rootg g^{rr} \phi'(r) \right].
\ee
Throughout this work, the prime denotes derivative with respect to $r$.
It will also be useful to express the second derivative of the potential as
\be
	\label{Udoubleprime}
	\frac{d^2 U(\phi(r))}{d\phi^2}= \frac{1}{\phi'(r)} \partial_{r} \left[\frac{dU(\phi(r))}{d\phi} \right].
\ee  

One can write the field $\phi(r)$ in terms of the metric components by noting that
\be
	g^{00}G_{00} - g^{rr}G_{rr} = F_{0}(r) - F_{r}(r)
\ee
and
\ba
	8 \pi G_{p+2} \left(g^{00} T_{00} -g^{rr} T_{rr} \right) &=&
	\thalf \left[
	- \mathcal{L}_\phi  - \left( g^{rr} \phi'(r)^2 -
	\mathcal{L}_\phi (r) \right) \right] 
	\\ &=& - \thalf g^{rr} \phi'(r)^2.
\ea
Thus, 
\be
	\label{phieqn} 
	\phi'(r)^2 = 2g_{rr}\left(F_0(r) - F_r(r)\right).
\ee

It is also noteworthy that this background has the special property $F_0 = F_x$ as can be seen by considering
\ba
	g^{00}G_{00} - g^{xx}G_{ii} &=& -8 \pi G_{p+2} \left(g^{00}T_{00} - g^{xx}T_{ii}\right)\\
	F_{0}(r) - F_x(r) &=& \mathcal{L}_\phi(r) - \mathcal{L}_\phi(r) = 0.  
	\label{F0Fx}
\ea
Because of this fact, and the general theorem given in \cite{BuchelLiu}, all backgrounds we consider have
$\eta/s = 1/4 \pi$.  

\section{Hydrodynamic Fluctuations}

We now introduce fluctuations of the fields on this background
$g_{\mu \nu} \rightarrow g_{\mu \nu} + \delta g_{\mu \nu}$ and $\phi
\rightarrow \phi + \delta \phi$, and assume the usual time dependence
\ba
	\delta g_{\mu \nu}(t,z,r) &=& e^{i(qz - wt)}h_{\mu \nu}(r), \\
	\delta \phi(t,z,r) &=& e^{i(qz - wt)} \delta \phi(r). 
\ea
Here we use the coordinate $z$ to denote one of the spatial
coordinates: $z \equiv x_p$, and $w$ and $q$ are the energy and
momentum of the perturbation.  For the sound mode, in the gauge where
$h_{\mu r} = 0$, the only non-zero fluctuations are \cite{hydroI, Mas}
\ba
	h_{00}(r) & \equiv & g_{00}(r)A(r), \label{Adef}\\
	\frac{1}{p-1}\sum_{i=1}^{p-1} h_{ii}(r) & \equiv & g_{xx}(r) B(r), \label{Bdef}\\
	h_{zz}(r) & \equiv & g_{xx}(r)C(r),\label{Cdef} \\
	h_{0z}(r) & \equiv & g_{00}(r)D(r),\label{Ddef} \\
	\delta \phi(r) \label{dphidef}.
\ea
Turning on these perturbations, and expanding the background equations
of motion to first order in the perturbation, we end up with 
a set of 8 differential equations for the perturbations.  The equations 
are presented in full in Appendix \ref{linearizedeqns}.  

\section{Gauge Invariant Equations}
There is still residual gauge freedom in these equations under the infinitesimal diffeomorphisms 
\ba
	h_{\mu \nu} &\rightarrow& h_{\mu \nu} - \nabla^{(0)}_\mu \xi_\nu - \nabla^{(0)}_\nu \xi_\mu, \\  
	(\delta \phi) &\rightarrow& (\delta \phi) - \xi_\mu (\partial^\mu \phi)
\ea
for any vector $\xi_\mu$ = $\xi_\mu(r)e^{i(qz-wt)}$.  The following 
gauge invariant combinations given in \cite{Mas} transform only into themselves
under such a diffeomorphism.  (i.e. $Z_0 \rightarrow Z_0$ and $Z_\phi \rightarrow Z_\phi$.)  
\be
	Z_0(r) = -f(r)^2 \left(q^2A(r) + 2qw D(r)\right) +w^2 C(r) 
	- \left( q^2 \frac{g_{00}'(r)}{g_{xx}'(r)}+w^2 \right)B(r) 
\ee
\be
	Z_\phi(r) = \delta \phi(r) - \frac{\phi'(r)}{\DL\left[g_{xx}(r)\right]}B(r)
\ee
By taking an appropriate combination of the
equations (\ref{g00eqn}-\ref{scalareqn}), one can arrive at two
coupled second order equations for the gauge invariant combinations.
The particular combination of the Einstein equations used, as well as additional 
details regarding this derivation, are presented in Appendix \ref{gaugeappendix}.  A combination which appears
frequently in these equations is
\be
	\alpha(r) \equiv q^2 \left((p-1)+ \frac{\DL [g_{00}(r)]}{\DL[g_{xx}(r)]}\right) - \frac{p w^2}{f(r)^2}.
\label{alphadef}
\ee

The following is the most compact form of the gauge invariant equations that
we have been able to find.
\ba
	\label{z0eqn}
	\frac{g_{rr}}{\rootg} \alpha^2 f^4 \partial_r \left[\frac{\rootg g^{rr}}{\alpha^2 f^4}Z_0'\right] 
	+Z_0 \left( \DL[f^2]\DL[f^2\alpha] - g_{rr}\left(w^2g^{00} + q^2g^{xx}\right)\right) &+& \\
	2Z_{\phi} \phi' f^2 \left( \alpha \partial_r \left[ \frac{1}{\alpha} \left(\frac{w^2}{f^2} -q^2 \right) \right] 
	+ \frac{q^2 \DL[f^2]}{p \DL[g_{xx}]} \DL \left[ \rootg g^{rr} \phi' \right] \right)
	&=&0 \nonumber
\ea
\ba
	\label{zphieqn}
	\frac{g_{rr}}{\rootg}  \partial_r \left[\rootg g^{rr}Z_\phi'\right] 
	+\frac{2}{\alpha} \partial_r \left[ \frac{\phi'}{\DL [g_{xx}]}\right] \left \{\frac{1}{f} \partial_r \left[\frac{Z_0}{f} \right] -
	\left(w^2 g^{00} + q^2 g^{xx} \right)\phi'g_{xx} Z_\phi \right\} &\,& \\
	-Z_\phi \left\{g_{rr}\left(q^2 g^{xx} + w^2 g^{00}\right) 
	+ \frac{\left(\DL [g_{xx}]\right)^2}{f^2 \phi'} \partial_r\left[\frac{f^2 \phi'}{\left(\DL [g_{xx}]\right)^2}\DL \left[ \rootg g^{rr} \phi' \right] \right] 
	\right\} &=&0 \nonumber
\ea
Of course, one can write these equations only in terms of the metric
components because of the relation
(\ref{phieqn}).  

To determine the dispersion relation, one needs to solve the above
equations perturbatively in $w,q$, and then apply the incoming wave
and Dirichlet boundary conditions \cite{Kovtun1}.  We will illustrate
this procedure in the following section.  Some terms can be neglected
when considering these equations to $\mathcal{O}(w,q)$, but one should
keep in mind that $Z_0$ is $\mathcal{O}(q^2)$, $Z_\phi$ is
$\mathcal{O}(1)$, and $\alpha$ is $\mathcal{O}(q^2)$.

\section{Example Application}

As a simple example of how to apply these equations, consider the
special case of $\phi(r) = C_1 \log \left[g_{xx}(r)\right]$ where
$C_1$ is a constant.  In this special case, the equation (\ref{zphieqn}) reduces to

\be
	\frac{g_{rr}}{\rootg}  \partial_r \left[\rootg g^{rr}Z_\phi'\right] 
	-Z_\phi g_{rr}\left(q^2 g^{xx} + w^2 g^{00}\right) = 0.
	\label{specialphi}
\ee  
This can be more easily seen by re-writing
\ba
	&\,& \partial_r\left[\frac{f^2 \phi'}{\left(\DL [g_{xx}]\right)^2}\DL \left[ \rootg g^{rr} \phi' \right] \right]\\ 
	&=&  C_1  \partial_r\left[\frac{f^2}{\DL[g_{xx}]} \DL \left[ \rootg g^{rr} \DL[g_{xx}] \right] \right] \\
	&=&  C_1 \left\{ \frac{2g_{rr}f^2}{\DL[g_{xx}]} (F_0 - F_x) - 
		f^2 \partial_r \left[ \frac{2 g_{rr}(F_0 -F_r)}{p \left(\DL[g_{xx}]\right)^2} \right] \right\}\\
	&=&  C_1 \left\{ \frac{2g_{rr}f^2}{\DL[g_{xx}]} (F_0 - F_x) - 
		f^2 \partial_r \left[ \frac{\phi'(r)^2}{p \left(\DL[g_{xx}]\right)^2} \right] \right\}\\
	&=& 0.  
\ea

The standard way of applying the incoming wave boundary conditions is to make the ansatz \cite{Kovtun1,Mas}
\ba
	Z_0(r) &=& f(r)^{-\frac{i w}{2 \pi T}} \left(Y_0(r)+ q Y_1(r)+ ... \right) \\
	Z_{\phi}(r) &=& f(r)^{-\frac{i w}{2 \pi T}} \left(Y_{\phi 0}(r)+ q Y_{\phi 1}(r) + ... \right)\\
	w(q) &=& w_1 q + w_2 q^2 + ...
	\label{ansatz}
\ea
with the condition that all $Y$ functions are regular at the horizon.
Inserting this ansatz into (\ref{specialphi}), expanding the result in
powers of q, and neglecting terms of $\mathcal{O}(q^2)$ and higher, we
find the following equation:

\be
	\partial_r \left[ \rootg g^{rr} Y_{\phi 0}' \right] 
	+ q\, \partial_r \left[ \rootg g^{rr} \left( Y_{\phi 1}' - \frac{i w_1}{2 \pi T} \DL(f) Y_{\phi 0} \right) \right]. 
\ee
Solving this order by order in $q$ is now quite simple.  The solution for $Y_{\phi 0}$ can be written in terms of an integral
\be 
	Y_{\phi 0}(r) = c_0 + c_1 \int_{r_0}^{\infty} \frac{g_{rr}(r')}{\sqrt{-g(r')}} dr',
\ee
but this integral is logarithmically divergent at the horizon by (\ref{NH}).  Thus,
the assumption of regularity on the $Y$ functions leads to $Y_{\phi 0}
= c_0$.  Finally, this constant must be set to zero by the Dirichlet
boundary condition at infinity.  Plugging $Y_{\phi 0} = 0$ into
the next order equation, one also finds $Y_{\phi 1} = 0$.

Next, one must solve the equation for $Z_0$ using these boundary
conditions with the knowledge that $Z_\phi = 0$.  But before doing so,
it is useful to pause and ask what type of metrics these results will
be applicable to.  There are 3 unknown metric functions, $g_{00},
g_{xx}, g_{rr}$, but we have two constraints on them, namely
(\ref{phieqn}) and (\ref{F0Fx}).  It can be shown using these constraints, 
and the assumption of the scalar field profile $\phi \sim \log(g_{xx})$ (as 
at the beginning of this section), that we are considering metrics which satisfy $F_0 = F_x$ and the following
constraint:
\be
	g_{00}(r) = \frac{a_0}{a_2-p} g_{xx}(r) + a_1 g_{xx}(r)^{a_2-p+1}.
	\label{specialmetric}
\ee 
Here, $a_0,a_1,a_2$ are independent of $r$, and we have defined the
coefficients as above for future convenience.  Clearly, $a_1$ can be
determined in terms of $a_0$ and $a_2$, by requiring $g_{00}$ vanish
at the horizon.

Returning now to the equation for $Z_0$, going through the same steps
of inserting the incoming wave ansatz, expanding in powers of $q$,
solving order by order in $q$, and applying the boundary conditions
leads to the dispersion relation
\be
	w(q) = \sqrt{\frac{a_0 - a_2}{p}}q - i \frac{a_2}{2 \pi T p} q^2 + \mathcal{O}(q^3). 
	\label{dispersionsolution}
\ee
The details of this derivation are presented in the Appendix \ref{specialcaseappendix}.
Comparing this dispersion relation with the expected hydrodynamic
dispersion relation (\ref{sounddispersion})
yields the relations for the speed of sound and bulk viscosity:
\ba
	v_s &=& \sqrt{\frac{a_0 - a_2}{p}} \nonumber\\
	\frac{\zeta}{\eta} &=& \frac{2(2 a_2 - p + 1)}{p}. 
	\label{transportsolution}
\ea

Let us now explicitly check these results agree with known cases.
Consider the `Schwarzschild AdS Black Hole' metric in p+2 dimensions,
\ba
	ds^2 &=& \frac{r^2}{L^2} \left[ -f(r)^2 dt^2 + dx_j dx^j \right] + \frac{L^2 dr^2}{r^2 f(r)^2}\\
	f(r)^2 &=& 1 - \left(\frac{r_0}{r} \right)^{p+1}
\ea
This is equivalent to the metric (\ref{specialmetric}) with the
choices $a_0 = (p+1)/2, a_2 = (p-1)/2$.  Inserting these
into (\ref{transportsolution}) yields
\ba
	v_s &=& 1/\sqrt{p} \nonumber\\
	\zeta &=& 0
\ea
which are in perfect agreement with our expectations, since the metric
is conformal, and is thus dual to a conformal field theory.

Next, consider the `Dp-Brane metric' in the Einstein frame, which can be reduced to \cite{stretched, Mas}
\ba
	ds^2 &=& \left(\frac{r}{L}\right)^\frac{9-p}{p}\left[-f(r)^2 dt^2+dx_jdx^j\right] 
	+ \left( \frac{r}{L} \right)^{\frac{p^2 - 8p +9}{p}}\frac{dr^2}{f(r)^2}, \\
	f(r)^2 &=& 1-\left(\frac{r_0}{r}\right)^{7-p}.  
\ea
This is the same as (\ref{specialmetric}) with the choices $a_0 =
(7p - p^2)/(9-p)$ and $a_2 = 2p/(9-p)$.  Inserting these into
(\ref{transportsolution}) gives
\ba
	v_s &=& \sqrt{\frac{5-p}{9-p}} \nonumber\\
	\frac{\zeta}{\eta} &=& \frac{2(3-p)^2}{9-p},
\ea
which is in agreement with the result of \cite{Mas}.  

It has been conjectured \cite{Buchel} that the relationship 
\be
	\frac{\zeta}{\eta} \geq 2\left( \frac{1}{p} - v_s^2 \right)
\ee
should hold for a strongly coupled plasma in $p$ spatial dimensions.  
From our formulas, one can see that this relation is satisfied provided
\be
 a_0 + a_2 \geq p.
\ee
Both special cases considered above have $a_0 + a_2 = p$, and thus
saturate the conjectured bound.  
At first sight it appears that it would be possible to violate the
conjectured bound with an appropriate choice of metric, but whether or
not such a gravity dual could be embedded into string theory, and
whether or not such a model would provide a reasonable description of
a strongly coupled plasma should perhaps be investigated in the
future.

This completes the example calculation for the single scalar field
profile chosen.  In the process of this example, we have generated
equation (\ref{transportsolution}), which is applicable to metrics
which obey $F_0 = F_x$, and (\ref{specialmetric}).  Equations (\ref{transportsolution})
are generalizations of formulas given in \cite{Mas}.

This is only one (particularly simple) application of the gauge
invariant equations in the previous section.  In the next section we
increase the generality of the gauge invariant equations by allowing
for multiple scalar fields.

\section{Multiple Scalar Fields}
Consider now adding additional scalar fields, so that the
total number is $n$.  
\be
	\mathcal{S} = \frac{1}{16 \pi G_{p+2}} \int \, d^{p+2}x \rootg 
	\left( R  - \thalf \sum_{k=1}^n \partial_\mu \phi_k \partial^\mu \phi_k - U(\phi_1,\phi_2...\phi_n) \right). 
\ee
There is now a background equation for each scalar field.  As before, assume each field only depends
on the radial coordinate $r$.  
\be 
	\Box \phi_k = \frac{1}{\rootg} \partial_r \left[\rootg g^{rr} \phi_k'(r) \right] = \frac{\partial U(\phi_1,\phi_2,...\phi_n)}{\partial \phi_k}. 
\ee
Adding these all together gives
\be
	\sum_{k=1}^n \Box \phi_k = \sum_{k=1}^n \frac{1}{\rootg} \partial_r \left[\rootg g^{rr} \phi_k'(r) \right] 
	= \sum_{k=1}^n \frac{\partial U(\phi_1,\phi_2...\phi_n)}{\partial \phi_k} 	  
\ee
In addition, the energy momentum tensor now becomes 
\ba
	8 \pi G_{p+2} T_{\mu \nu} &=& \thalf \sum_{k=1}^n \left(\partial_\mu \phi_k \partial_\nu \phi_k - g_{\mu \nu} \mathcal{L}_{\phi k} \right) \\
	\mathcal{L}_{\phi k} &=& \thalf \sum_{k=1}^n \partial_\lambda \phi_k \partial^\lambda \phi_k + U(\phi_1,\phi_2...\phi_n)
\ea
and the constraint equation (\ref{phieqn}) becomes
\be
	\label{phikeqn} 
	\sum_{k=1}^n \phi_k'(r)^2 = 2g_{rr}\left(F_0(r) - F_r(r)\right).
\ee

It is fairly straightforward to generalize the gauge invariant equations to include more scalar fields.
The linearized Einstein equations are modified due to the presence of other fields, and 
one needs to introduce an additional gauge invariant variable for each additional scalar field.  We use
the index $i$ to denote a particular scalar field with $i = 1, 2...n$.   
\be
	Z_{\phi i}(r) = \delta \phi_i(r) - \frac{\phi_i'(r)}{\DL\left[g_{xx}(r)\right]}B(r)
\ee
One then gets additional terms in the differential equation for $Z_0$.  It can be written:
\ba
	\label{multiscalarz0}
	\frac{g_{rr}}{\rootg} \alpha^2 f^4 \partial_r \left[\frac{\rootg g^{rr}}{\alpha^2 f^4}Z_0'\right] 
	+ Z_0 \left( \DL[f^2]\DL[f^2\alpha] - g_{rr}\left(w^2g^{00} + q^2g^{xx}\right)\right) &+& \nonumber \\
	\sum_{k=1}^n 
	\left\{ 
		2Z_{\phi k} \phi_k' f^2 \left( \alpha \partial_r \left[ \frac{1}{\alpha} \left(\frac{w^2}{f^2} -q^2 \right) \right]
		+ \frac {q^2 \DL[f^2]}{p \DL[g_{xx}]} \DL \left[ \rootg g^{rr} \phi_k' \right] \right)
	\right\} &=&0
\ea
There are also $n$ gauge invariant equations, one for each of the scalar
fields.  Each of these equations takes the form
\ba
	&\,& \frac{g_{rr}}{\rootg}  \partial_r \left[\rootg g^{rr}Z_{\phi i}'\right] 
	-Z_{\phi i} g_{rr}\left(w^2 g^{00} + q^2 g^{xx}\right)\nonumber -g_{rr}\sum_{k=1}^n Z_{\phi k} \frac{\partial^2 U}{\partial \phi_i \partial \phi_k}\\
	&-& \frac{2\phi_i'}{p \DL [g_{xx}] \alpha} \left\{ \sum_{k=1}^n \left[ Z_{\phi k} \phi_k' \left(\alpha \DL \left[\rootg g^{rr} \phi_k' \right]
	  + p g_{xx} \DL \left[\frac{\phi_i'}{\DL [g_{xx}]}\right]\left(w^2 g^{00} + q^2 g^{xx}\right) \right) \right] \right\} \nonumber \\
	&+& \frac{2}{\alpha f} \partial_r \left[ \frac{\phi_i'}{\DL[g_{xx}]} \right] \partial_r \left[ \frac{Z_0}{f} \right] = 0.
\label{multiscalarzphi}
\ea At first sight, this equation appears slightly different than that
of a single scalar field (\ref{zphieqn}).  There are two reasons for
this.  First, in the case of a single scalar field, the term $
\frac{\partial^2 U}{\partial \phi_i \partial \phi_k}$ can be written
in terms of the metric only due to (\ref{Udoubleprime}); one cannot do
this if the potential depends on more than one variable.  Secondly, in
the case of a single scalar field we removed a term which vanishes by
the equations of motion, but we have not done so here.  The 
term in question appears in (\ref{Z2eqn}).

\section{Conclusion}
In this work we have presented a set of gauge invariant equations for
sound mode perturbations on a generic black brane type background.
The equations (\ref{multiscalarz0}-\ref{multiscalarzphi}) are the
main results of this paper.  These equations can be used to determine
the speed of sound and bulk viscosity for any metric which can be
generated by a set of minimally coupled scalar fields.
In order to determine the dispersion relation, one must solve these
equations perturbatively in $q$, applying the incoming wave boundary
condition at the horizon, and Dirichlet boundary condition at $r =
\infty$.  

The gauge invariant equations are quite complicated, and so far a
general analytic solution has eluded us, though we did present a
solution for a particular class of metrics.  Metrics which obey
(\ref{specialmetric}) and $F_0 = F_x$  have speed of sound and bulk viscosity given by
(\ref{transportsolution}).  These results are a generalization of the
results of \cite{Mas}, and include both Dp-brane, and Schwarzschild AdS
black hole metrics.

In the future, we hope to report further on the possibility of an
analytic solution of these equations.  In the event such a solution is
not possible, the equations can be solved 
numerically for a specified metric and set of scalar field profiles.
One could also increase the generality of the gauge invariant
equations by including other types of matter such as gauge fields.

We believe that the equations derived here may be useful for
phenomenologically based models such as those of
\cite{Softwall1,Softwall2}.  It would be interesting to see what these
models (which were chiefly developed to match the meson mass
spectrum) have to say about hydrodynamics.

\section*{Acknowledgments}
I would like to thank Joe Kapusta for helpful comments.  I would also
like to thank the Physics Department at McGill University, and the
organizers of the workshop ``AdS/CFT, Condensed Matter, and QCD'' for
their hospitality.  This work was supported by the US Department of
Energy (DOE) under Grant. NO. DE-FG02-87ER40328, and by the Graduate
School at the University of Minnesota under the Doctoral Dissertation
Fellowship.

\appendix
\section{Background Ricci Tensor}
\label{ricciappendix}

An explicit computation of the Ricci tensor for our background gives
\ba
	F_0(r) &\equiv& g^{00}R_{00} = \frac{1}{2\rootg} \partial_r \left( \rootg g^{rr} \DL[g_{00}] \right), \\
	F_x(r) &\equiv& g^{ii}R_{ii} = \frac{1}{2\rootg} \partial_r \left( \rootg g^{rr} \DL[g_{xx}] \right), \\
	F_r(r) &\equiv& g^{rr}R_{rr} = \frac{1}{4 g_{00}^{\,\prime}}\partial_r \left(g_{00}g^{rr}\DL[g_{00}]^2 \right)
	  +  \frac{p}{4 g_{xx}^{\, \prime}}\partial_r \left(g_{xx}g^{rr}\DL[g_{xx}]^2 \right). 
\ea
Here $i$ denotes the spatial coordinates $x_1...x_p$, and we are using the notation $\DL$ to denote the logarithmic
derivative as defined in the text (\ref{DL}).

Combinations which appear frequently in the text are 
\be
	F_0 - F_x = \frac{1}{\rootg} \partial_r \left( \rootg g^{rr} \DL[f] \right),  
\ee
and 
\be
	F_0 - F_r = \frac{p}{2}g^{rr} \DL[g_{xx}] \DL \left[ \frac{f \sqrt{g_{rr}}}{\DL[g_{xx}]} \right],  
\ee  
where $f$ is defined as in the text (\ref{fdef}).  

\section{Linearized Perturbation Equations}
\label{linearizedeqns}

Hydrodynamic behavior of the dual gauge theory is accessed by
introducing perturbations on top of the black brane background.
Turning on the sound mode perturbations (\ref{Adef}-\ref{dphidef}),
and expanding the background equations of motion to first order in the
perturbation results in the following set of equations.  We
use the superscript ``1'' to denote quantities that are $\mathcal{O}(h_{\mu
\nu})$ or $\mathcal{O}(\delta \phi)$.
\begin{center}$G^{(1)}_{00} = -8 \pi G_{p+2} T^{(1)}_{00}$ 
\ba
	\label{g00eqn}
	\frac{g_{rr}}{\rootg}f \partial_r \left[\frac{\rootg g^{rr}}{f} ((p-1)B'+C') \right]
	+g_{rr}\frac{dU}{d\phi}(\delta \phi) + \phi'(\delta \phi)'-(p-1) q^2 g_{rr}g^{xx} B=0 \\
	\nonumber
\ea
$\displaystyle\sum_{i=1}^{p-1}G^{(1)}_{ii} = \displaystyle\sum_{i=1}^{p-1}\left( -8 \pi G_{p+2} T^{(1)}_{ii}\right)$
\ba	
	\label{gxxeqn}
	\frac{g_{rr}}{f \rootg}\partial_r \left[\rootg g^{rr} f A' \right] 
	+ \frac{g_{rr}}{\rootg}\partial_r \left[\rootg g^{rr}\left((p-2)B'+C'\right) \right]
	+g_{rr}\frac{dU}{d\phi}(\delta \phi) + \phi'(\delta \phi)' &\,&  \\
	- g_{rr}\left(q^2g^{xx}\left(A+(p-2)B\right) + w^2 g^{00}\left((p-2)B+C\right)+2qwg^{xx}D\right) &=&0 \nonumber\\
	\nonumber 
\ea
$G^{(1)}_{zz} = - 8 \pi G_{p+2} T^{(1)}_{zz} $
\ba
	\label{gzzeqn}
	\frac{g_{rr}}{f \rootg}\partial_r \left[\rootg g^{rr} f A' \right] 
	+ \frac{g_{rr}}{\rootg}\partial_r \left[\rootg g^{rr}(p-1)B' \right] 
	+ g_{rr}\frac{dU}{d\phi}(\delta \phi) + \phi'(\delta \phi)' &\,& \nonumber\\
	- (p-1) w^2g_{rr}g^{00}B &=&0\\
	\nonumber 
\ea
$G^{(1)}_{0z} = -8 \pi G_{p+2} T^{(1)}_{0z} $
\ba
	\label{g0zeqn}
	 \frac{\rootg (g^{xx})^{p+1}}{f^2} \partial_r \left[ \frac{(g_{xx})^{p+1}}{\rootg} \partial_r \left(f^2 D \right)\right] 
	-2g_{rr}(F_0 - F_x)D + q w (p-1) g_{rr}g^{00}B =0\\
	\nonumber 
\ea
$G^{(1)}_{rr} = -8 \pi G_{p+2} T^{(1)}_{rr} $
\ba
	\label{grreqn}
	\DL \left[(g_{xx})^p\right]A' + \DL \left[g_{00}(g_{xx})^{p-1}\right]((p-1)B'+C')
	-2\phi'(\delta \phi)'+2g_{rr}\frac{dU}{d\phi}(\delta\phi) &\,& \nonumber \\
	-2g_{rr}\left(w^2g^{00}((p-1)B+C) + q^2 g^{xx}(A+(p-1)B) +2qwg^{xx}D\right) &=&0\\
	\nonumber
\ea
$G^{(1)}_{0r} = -8 \pi G_{p+2} T^{(1)}_{0r}$ 
\ba
	\label{g0reqn}
	w f \partial_r\left[ \frac{1}{f} ((p-1)B+C) \right] -q f^2 D' + w\phi'(\delta \phi) =0\\
	\nonumber 
\ea
$G^{(1)}_{rz} = - 8 \pi G_{p+2} T^{(1)}_{rz}$
\ba
	\label{grzeqn}
	\frac{q}{f}\partial_r \left[ f A \right] + \frac{w}{f^2} \partial_r \left[f^2 D\right]+ q (p-1)B'+ q \phi'(\delta \phi) =0\\
	\nonumber 
\ea
$\Box^{(0)}(\delta \phi) + \Box^{(1)}\phi = \delta \left(\frac{dU(\phi)}{d\phi}\right)$
\ba
	\label{scalareqn}
	\frac{g_{rr}}{\rootg} \partial_r \left[\rootg g^{rr} (\delta \phi)' \right] + \thalf \phi'H'
	-g_{rr} \left(w^2 g^{00} + q^2 g^{xx}+\frac{d^2 U}{d\phi^2}\right)(\delta \phi) &=&0
\ea
\end{center}
where we have defined
\be
	H(r)  \equiv   A(r) + (p-1)B(r) + C(r). 
\ee

\section{Derivation of Gauge Invariant Equations}
\label{gaugeappendix}
Here we present the explicit combination of the Einstein equations
that lead to the gauge invariant equations.  Consider the following 
combination of the Einstein equations (\ref{g00eqn} - \ref{scalareqn}): 
\ba
	&f^2& \left\{ \frac{\alpha}{p} \left[(\ref{g00eqn}) - (\ref{gxxeqn}) + (\ref{gzzeqn}) + \thalf (\ref{grreqn}) \right] 
	  + \left[(\ref{g00eqn}) + \thalf (\ref{grreqn})\right] \left(\frac{w^2}{f^2}\right) - 2 q w (\ref{g0zeqn}) 
	  \right. \nonumber \\
	  &-& q^2 \Biggl. \left[\thalf (\ref{grreqn})+(\ref{gzzeqn}) \right] \Biggr\}
	 + 2 \left \{ q f^2 \DL[\alpha] (\ref{grzeqn}) - w \DL[\alpha f^2 ] (\ref{g0reqn}) \right\}.  
\ea
Here $\alpha$ is defined as in the text (\ref{alphadef}).  
After a long calculation, one can show that this reduces to
\be
	\mathcal{Z}_1 - 2 g_{rr}(F_0 - F_x)\left(w^2(B-C)+Z_0\right) - B q^2 f^2 \Delta_1 = 0,
\ee
where $\mathcal{Z}_1$ is the left side of (\ref{z0eqn}), and 
\ba
	\Delta_1 &=& 2 \DL[g_{xx}]\alpha^2 \partial_r \left[ \frac{g_{rr}}{\alpha^2 \DL[g_{xx}]^2}(F_0 - F_x) \right] 
	- \frac{4 g_{rr}}{p \DL[g_{xx}]^2}(F_0 - F_x) (\phi')^2 \nonumber \\  
	&+& \alpha^2 \DL[f^2] \partial_r \left[\frac{1}{p \alpha^2 \DL[g_{xx}]^2} \left(2 g_{rr}(F_0 - F_r) - (\phi')^2\right) \right].
\ea
Clearly, $\Delta_1$ vanishes by the background equations of motion (\ref{F0Fx}) and (\ref{phieqn}), leaving 
a differential equation involving only the gauge invariant variables, namely, (\ref{z0eqn}).

To derive the other gauge invariant equation, the relevant combination is:
\be
	(\ref{scalareqn})
	+ \frac{\phi'}{p \DL[g_{xx}]}\left\{ (\ref{gxxeqn}) - (\ref{gzzeqn}) - (\ref{g00eqn}) - \frac{1}{2} (\ref{grreqn}) \right\} 
	+ \frac{2}{\alpha} \partial_r \left[ \frac{\phi'}{\DL[g_{xx}]} \right] \left\{ \frac{w}{f^2}(\ref{g0reqn}) - q (\ref{grzeqn}) \right\}.  
\label{scalarcombo}
\ee
Another lengthy calculation reduces this combination to
\be
	\mathcal{Z}_2 - \frac{2}{p \DL[g_{xx}]} \left((\phi')^2 - 2 g_{rr}(F_0-F_r)\right)\DL \left[ \rootg g^{rr} \phi' \right]Z_{\phi} - B \Delta_2 = 0,
\label{Z2eqn}
\ee
where $\mathcal{Z}_2$ is the left side of (\ref{zphieqn}), and 
\ba
	\Delta_2 &=& \frac{\phi'}{p \DL[g_{xx}]^2} \left(\frac{g_{rr}}{\rootg} \right)^2 \partial_r \left[\left(\rootg g^{rr} \right)^2((\phi')^2 - 2g_{rr}(F_0 - F_r)) \right] \nonumber \\
	&+& \frac{2}{\alpha \DL[g_{xx}]} \partial_r \left[ \frac{\phi'}{\DL[g_{xx}]}\right] \left((\phi')^2 - 2g_{rr}(F_0-F_r)\right)\left(q^2 - \frac{w^2}{f^2} \right) \nonumber \\
	&+& \frac{2 g_{rr}}{\DL[g_{xx}]}\left( \frac{2 q^2}{\alpha} \partial_r \left[ \frac{\phi'}{\DL[g_{xx}]}\right]+\phi' \right)(F_0 - F_x). 
\ea
Again, all terms except for $\mathcal{Z}_2$ vanish by the background equations of motion, leaving us with (\ref{zphieqn}).  

Generalization of these equations to multiple scalar fields is quite
straightforward.  In this case, the Einstein equations (\ref{g00eqn} -
\ref{scalareqn}) are modified, but the combinations that lead to the
gauge invariant equations remain unchanged.  As mentioned in the text,
there will be an additional gauge invariant equation for each
additional scalar field.  In this case, there will be $n$ combinations
like (\ref{scalarcombo}).  Each will have (\ref{scalareqn}) replaced
by the analogous equation for each particular scalar field.

\section{Special Case Solution}
\label{specialcaseappendix}
In this section, we present the remainder of the calculation which leads to equations
(\ref{transportsolution}).  One must go back to (\ref{z0eqn}), insert $Z_{\phi} = 0$ and
the incoming wave ansatz (\ref{ansatz}), and expand the resulting equation in powers of $q$.  
In doing so, one should take into account the constraints (\ref{F0Fx}) and (\ref{specialmetric}).  
The first of these constraints allows one to eliminate $\rootg g^{rr}$ in favor of $\DL[f^2]$.  
Once this is done, the lowest order equation for $Y_0$ can be written
\be
	\frac{\left(C_0 - f^2 a_2\right)^2}{a_2 f^2} \partial_r \left[ \frac{1}{\DL[f^2]\left(C_0 - f^2 a_2 \right)^2}Y_0' \right]
	- \frac{\DL[f^2]}{C_0 - f^2 a_2} Y_0 = 0,
\ee  
where 
\be
	C_0 = p w_1^2 - a_0. 
\ee
The general solution to that equation contains two arbitrary constants $k_1$ and $k_2$, and can be written
\be
	Y_0(r) = (p w_1^2- a_0 + f(r)^2 a_2)\left[k_0 
	+ k_1 \int_{r}^{\infty}\, \left( \frac{p w_1^2 - a_0 - a_2 f(r')^2}{p w_1^2 - a_0 + a_2 f(r')^2}\right)^2 \frac{f'(r')}{f(r')} \, dr' \right],
\label{Y0soln}
\ee
as can be found by first guessing a solution $Y_0(r) = b_0 + b_1
f(r)^2$, and then using the technique of reduction of order once this
solution is found.

The integral in (\ref{Y0soln}) is logarithmically divergent near the horizon, and thus the assumption of 
regularity leads to $k_1 = 0$.  Finally, applying the Dirichlet boundary condition at $r \rightarrow \infty$, 
leads to 
\be
	p w_1^2 - a_0 + a_2 = 0,
\label{w1soln}
\ee  
where we have assumed that $f(r\rightarrow \infty) = 1$.  Thus, we find
the lowest order term in (\ref{dispersionsolution}).  

Proceeding now to the next order in $q$, and substituting the solutions for $w_1$ and $Y_0$, one finds the
following differential equation for $Y_1$:
\be
	\partial_r \left[ \frac{ Y_1'}{\DL[f^2](1+f^2)^2} \right] 
	+ \frac{2 f f'}{(1+f^2)^3}\left(Y_1 - w_1 k_0 \left( \frac{i a_2}{\pi T} + 2 p w_2 \right) \right) = 0.
\ee
The solution to the homogeneous part can be found using the same techniques as for the solution for $Y_0$ listed
above.  A particular solution to the inhomogeneous equation is obviously $Y_1(r)=\, $Constant.  This leads to the
general solution,
\be
	Y_1(r) = (f(r)^2 - 1)\left(k_2 + k_3 \int_{r}^{\infty} \frac{(f(r')^2+1)^2}{(f(r')^2-1)^2} \frac{f'(r')}{f(r')} \, dr' \right)
		+ w_1 k_0 \left( \frac{i a_2}{\pi T} + 2 p w_2 \right).  
\ee
The integral above, is again logarithmically divergent, leading to the requirement that $k_3=0$.  Applying now the
Dirichlet boundary condition at $r \rightarrow \infty$ leads to
\be
	w_1 k_0 \left( \frac{i a_2}{\pi T} + 2 p w_2 \right) = 0.
\label{w2soln}
\ee
Equation (\ref{dispersionsolution}) immediately follows from (\ref{w1soln}) and (\ref{w2soln}).  

\begin{thebibliography}{99}

   \bibitem{whitepapers}
    I. Aresene, \emph{et al.} (BRAHMS Collaboration), Nucl. Phys. {\bf A757}, 1 (2005); 
    B.B. Back, \emph{et al.} (PHOBOS Collaboration), Nucl. Phys. {\bf A757}, 28 (2005);
    J. Adams, \emph{et al.} (STAR Collaboration), Nucl. Phys. A {\bf A757}, 102 (2005);
    K. Adcox, \emph{et al.} (PHENIX Collaboration), Nucl. Phys. A {\bf A757}, 184 (2005);

    \bibitem{Molnar}
      D. Moln\'{a}r, M. Gyulassy, Nucl. Phys. {\bf A697}, 495 (2002), \emph{erratum - ibid} {\bf A703}, 893 (2002).    

    \bibitem{Huovinen}
      P. Huovinen, P.F. Kolb, U.W. Heinz, P.V. Ruuskanen, S.A. Voloshin, Phys. Lett {\bf B503}, 58 (2001);
      P. Huovinen in \emph{Quark-Gluon Plasma 3} eds. R.C. Hwa and X.N. Wang, World Scientific, Singapore (2004).  

   \bibitem{hydroreview}
   D.T. Son, A.O. Starinets, Ann.Rev.Nucl.Part.Sci. \textbf{57}, 95-118 (2007).

   \bibitem{Kovtun2}
   P. Kovtun and L. G. Yaffe, Phys. Rev. D {\bf 68}, 025007 (2003).

   \bibitem{Maldacena}
   J. Maldacena, Adv. Theor. Math. Phys. {\bf 2}, 231 (1998).
   
   \bibitem{Witten}
   E. Witten, Adv. Theor. Math. Phys.  {\bf 2}, 505 (1998).

   \bibitem{Gubser}  
   S.S. Gubser, I.R. Klebanov, A.M. Polyakov, Phys. Lett. B {\bf 428}, 105 (1998)

   \bibitem{Policastro1}
   G. Policastro, D. T. Son and A. O. Starinets, Phys. Rev. Lett. {\bf 87}, 081601 (2001).

   \bibitem{recipe}
   D. T. Son and A. O. Starinets, JHEP {\bf 09}, 042 (2002).

   \bibitem{hydroI}
   G. Policastro, D. T. Son and A. O. Starinets, JHEP {\bf 09}, 043 (2002).

   \bibitem{hydroII}
   G. Policastro, D. T. Son and A. O. Starinets, JHEP {\bf 12}, 054 (2002).
  
   \bibitem{Herzog1}
   C. P. Herzog, JHEP {\bf 0212}, 26 (2002).

   \bibitem{Herzog2} 
   C. P. Herzog, Phys. Rev. D {\bf 68}, 024013 (2003).

   \bibitem{Kovtun1}
   P.K. Kovtun and A. O. Starinets, Phys. Rev. D {\bf 72}, 086009 (2005).
	
   \bibitem{Mas}
   J. Mas and J. Tarr\'{i}o, JHEP {\bf 0705}, 036 (2007).

   \bibitem{stretched}
   P. Kovtun, D. T. Son and A. O. Starinets, JHEP {\bf 10}, 064 (2003). 
  
   \bibitem{Saremi}
   O. Saremi, arXiv:hep-th/0703170  

   \bibitem{Kapusta}
   J.I. Kapusta, T. Springer, Phys. Rev. D {\bf 78}, 066017 (2008)

   \bibitem{Natsuume}
   M. Natsuume and T. Okamura, Phys. Rev. D {\bf 77}, 066014 (2008).

   \bibitem{Liu}
   N. Iqbal, H. Liu, arXiv:0809.3808 [hep-th] 

   \bibitem{StarinetsMembrane}
   A. O. Starinets, arXiv:0806.3797 [hep-th]

   \bibitem{Kharzeev1}
   D. Kharzeev, K. Tuchin, JHEP {\bf 0809}, 093, (2008). 

   \bibitem{Kharzeev2}
   F. Karsch, D. Kharzeev, K. Tuchin,  Phys. Lett. B {\bf 663}, 217 ,(2008)
 
   \bibitem{Gubser1}
   S. S. Gubser, A. Nellore, S. S. Pufu, F.D. Rocha, 
   arXiv:0804.1950 [hep-th] 

   \bibitem{Gubser2}
   S. S. Gubser, S. S. Pufu, F. D. Rocha, JHEP {\bf 0808}, 085 (2008).  

   \bibitem{Weinberg}
   S. Weinberg, {\it Gravitation and Cosmology: Principles and Applications of the General Theory of Relativity}, 
   Wiley \& Sons, New York, 1972.

   \bibitem{BuchelLiu}
     A. Buchel and J. T. Liu, Phys. Rev. Lett. {\bf 93}, 090602 (2004).

   \bibitem{Buchel}
     A. Buchel, Phys. Lett. B {\bf 663}, 286 (2008).  

   \bibitem{Softwall1}
   B. Batell, T. Gherghetta, Phys.Rev.D {\bf 78}, 026002 (2008).  

   \bibitem{Softwall2}
   W. de Paula, T. Frederico, H. Forkel, M. Beyer, arXiv:0806.3830 [hep-ph] 
 
 

\end{thebibliography}
\end{document}